# Static Quark Potentials in Quantum Gravity

Wolfgang Beirl[a], Bernd A. Berg[b,1],
Balasubramanian Krishnan[a,2], Harald Markum[a] and
Jürgen Riedler[a]

[a] *Institut für Kernphysik, Technische Universität Wien
A-1040 Vienna, Austria*

[b] *Department of Physics and SCRI, Florida State University
Tallahassee, FL 32306, USA*

**Abstract**

We present potentials between static charges from simulations of quantum gravity coupled to an SU(2) gauge field on $6^3 \times 4$ and $8^3 \times 4$ simplicial lattices. The action consists of the gravitational term given by Regge's discrete version of the Euclidean Einstein action and a gauge term given by the Wilson action, with coupling constants $m_p^2$ and $\beta$ respectively. In the well-defined phase of the gravity sector where geometrical expectation values are stable, we study the correlations of Polyakov loops and extract the corresponding potentials between a source and sink separated by a distance $R$. We compare potentials on a flat simplicial lattice with those on a fluctuating Regge skeleton. In the confined phase, the potential has a linear form while in the deconfined phase, a screened Coulombic behavior is found. Our results indicate that quantum gravitational effects do not destroy confinement due to non-abelian gauge fields.

## 1 Introduction

The goal of modern day theoretical physics is to construct a grand unified theory of all fundamental forces known to man. A main problem in this endeavor is to include the gravitational interactions because one encounters severe difficulties in quantizing gravity. The perturbative treatment suffers from the well-known problems of the unboundedness of the action and non-renormalizability

[1] Work supported in part by DOE under Contracts DE-FG05-87ER40319 and DE-FC05-85ER2500.
[2] Lise-Meitner Postdoctoral Research Fellow sponsored by FWF under Project M078-PHY.

Preprint submitted to Elsevier Science          10 February 1995

[1]. Regge Calculus [2] provides a non-perturbative way for investigations of the Euclidean Einstein action, called Regge-Einstein action in this context, on a lattice and offers the possibility to construct a unified theory by coupling gauge fields to the skeleton. One of the interesting results of such investigations was the discovery of an "entropy dominated" smooth phase in quantum gravity, where the expectation values with respect to the pure Regge-Einstein action are stable [3–5]. The next logical question was to address the physical relevance of this regime. Efforts in this direction have been made by coupling a non-abelian gauge field to gravity [6]. If one assumes that the world without gravity is described by a grand unified asymptotically free theory, these numerical studies investigate the relation of the hadronic scale to the Planck scale. In particular, one is interested whether confinement exists and hadronic masses (deconfinement temperature, string tension, etc.) may be chosen small compared to the Planck mass. In this work, we perform a non-perturbative, finite temperature study of the behavior of the static quark potentials in the coupled system of quantum gravity and SU(2) gauge fields in four spacetime dimensions. We extract a value for the string tension in the confined phase and the Debye screening mass in the deconfined phase. We also measure potentials on the flat simplicial lattice and compare them with those obtained on the coupled system.

## 2 The Model

In Regge Calculus the edge lengths are taken to be the dynamical variables of the discretized spacetime manifold. In four dimensions, the geometry is Euclidean inside a pentahedron and the curvature is concentrated at the triangles. The quantization proceeds via the path integral with the Regge-Einstein action for pure gravity. The choice of the measure is an unresolved issue and investigations with several measures could not favor any of them [5] with a slight preference for the scale-invariant measure [7]. We choose a hypercubic triangulation with $N_s^3 \times N_t$ vertices and a scale-invariant measure in our simulations:

$$D[\{l^2\}] = \prod_l \frac{dl^2}{l^2}, \qquad (1)$$

where $l$ is used to denote the link label as well as the length. The addition of the SU(2) gauge term is straightforward and follows conventional lattice gauge theory by assigning SU(2) matrices to the links. The elementary plaquettes become triangles on the Regge skeleton. The system of SU(2) gauge fields coupled to quantum gravity has the action [6]

$$S = 2m_p^2 \sum_t A_t \alpha_t - \frac{\beta}{2} \sum_t W_t \, \text{Re}[\text{Tr}(1 - U_t)] \,, \qquad (2)$$

with the bare Planck mass $m_p$ and the inverse gauge coupling $\beta$. $A_t$ and $\alpha_t$ correspond to the area and the deficit angle of the triangle $t$. The weights $W_t$ describe the coupling of gravity to the gauge field. $U_t$ is the ordered product of SU(2) matrices around $t$. In contrast to the flat lattice, the unit matrix 1 in the action is important because the weight factors are dynamical. They are constructed such that the correct continuum limit is ensured in the limit of vanishing lattice spacing [8].

The Polyakov loop $P(R)$ in the short extent $L_t$ of the lattice describes the propagation of a heavy quark and acts as an order parameter. We introduce a quark source and a sink separated by a distance $R$ and calculate the correlations of the Polyakov loops at these points. As in conventional SU(2) lattice gauge theory at finite temperature $T$ [9,10], we extract from the correlation function

$$\langle P(0) P^\dagger(R) \rangle = \exp[-\frac{1}{T} V(R)], \qquad (3)$$

the quantity $V(R)$ corresponding to the potential between the heavy quark-antiquark pair. In the confinement phase, $V(R)$ should grow linearly for large $R$ due to an infinite free energy of isolated quarks, while in the deconfinement phase, one expects a screened Coulombic behavior:

$$\begin{aligned} V(R) &= \frac{-\alpha}{R} + \sigma R + C \qquad \text{(confinement)} \\ &= \frac{-\alpha}{R} \exp(-\mu R) + C \quad \text{(deconfinement)}, \end{aligned} \qquad (4)$$

where $\sigma$ is the string tension, $\alpha$ the Coulomb parameter, $\mu$ the Debye screening mass and $C$ a constant.

We define a length scale by setting the expectation value of the volume of a pentahedron $\langle v \rangle$ to be a constant. The quantity $l_0 = \langle v \rangle^{\frac{1}{4}}$ serves as a length unit. In the well-defined phase of gravity expectation values are stable and the physical Planck mass $m_P$ can be related to $l_0$ by $m_P = m_p \times l_0^{-1}$. For the hadronic masses $m_h$, one expects an asymptotic scaling law from the renormalization group equation [6]. If this scaling could be observed, our simulations relate to the physical region in the same way as standard lattice gauge calculations do. In addition, classical gravity should be recovered for macroscopic distances.

In the context of quantum gravity, there exist reasons to consider a fundamental length [11] as physical, which then defines a natural cutoff. In a natural regularization scheme, no renormalization would be necessary and $l_0$ would serve as the fundamental length unit. Problems with locality, expected to occur at the Planck length $l_P$ might be avoided by this approach. It would be attractive if Regge Calculus could be utilized to define a natural regularization scheme. Assuming that this is the case, the desired ratio of the Planck mass with a hadronic mass $m_p/m_h \approx 10^{18}$ is obtained for a certain finite value $\beta_P$. Ratios $m_p/T_c(\beta)$ that differ from the physical value imply that we are away from $\beta_P$ and $l_P$, just as in usual lattice gauge theory where one is away from the continuum limit $\beta \to \infty$ and $a \to 0$.

To address the issue of the two different length scales, the Planck mass should turn out large in comparison to the hadronic mass. In our finite temperature $L_s^3 \times L_t$ lattices, decreasing the length unit $l_0$ means increasing the inverse gauge coupling which is restricted by a critical value, $\beta < \beta_c$. To allow larger $\beta_c$ values the temporal extension has to be enlarged, making sure simultaneously that the smooth phase of gravity does not break down, $m_p^2 < m_c^2$, with the border of the smooth phase $m_c^2 > 0$. In this study, we extend the previous $N_t = 2$ investigation [6] to $N_t = 4$. As in pure SU(2) gauge theory, $\beta_c(N_t = 2) < \beta_c(N_t = 4)$ also turns out on the fluctuating skeleton. We find that $m_c^2 \approx 0.025$ is almost unaffected by the accompanying increase in $\beta_c$. The corresponding deconfining temperature is defined by $T_c = 1/(N_t \langle x_l \rangle_t)$ with $\langle x_l \rangle_t$ the average link length in the short direction of the skeleton. This gives from $N_t = 2$ to $N_t = 4$, an increase in the mass ratio $m_c/T_c$ by about a factor two.

## 3 Numerical Results

We performed our simulations on a $6^3 \times 4$ lattice first and then, to extract a value of the string tension, on an $8^3 \times 4$ lattice [12]. For the coupled system, we used about twenty thousand sweeps for equilibration. Our data rely on 66000 measurements for the $6^3 \times 4$ lattice and on 30000 for the $8^3 \times 4$ system. To compare our results with the flat simplicial lattice, we used for the pure gauge system several thousand equilibration sweeps with 25000 measurements on the $6^3 \times 4$ lattice and 30000 on the $8^3 \times 4$ system.

We obtain the potentials between external sources by computing the correlations of Polyakov loops along the main axes. Fig. 1(a) presents the data points for the static quark potentials on the smaller lattice in the presence of gravity, while Fig. 1(b) depicts the situation with gravity switched off. To the best of our knowledge, Fig. 1(b) is the first investigation of SU(2) potentials on a flat simplicial lattice. Previous calculations were done on hypercubic [10] and body-centered hypercubic [13] lattices. Fig. 1(a) shows that $V(R)$ with $m_p^2 = 0.005$ in the "entropy dominated" phase behaves very differently in the two phases of the gauge field. For $\beta = 1.55$ the potential rises steeply with $R$, while for $\beta = 1.60$ it is relatively flat suggestive of a screened Coulombic behavior. In Fig. 1(b) we compare with the potentials from pure gauge simulations on the flat simplicial lattice at two values of $\beta$ below and above the deconfinement transition ($\beta = 1.02$ and $\beta = 1.10$). Qualitatively, the shape of the potentials is similar to that of Fig. 1(a).

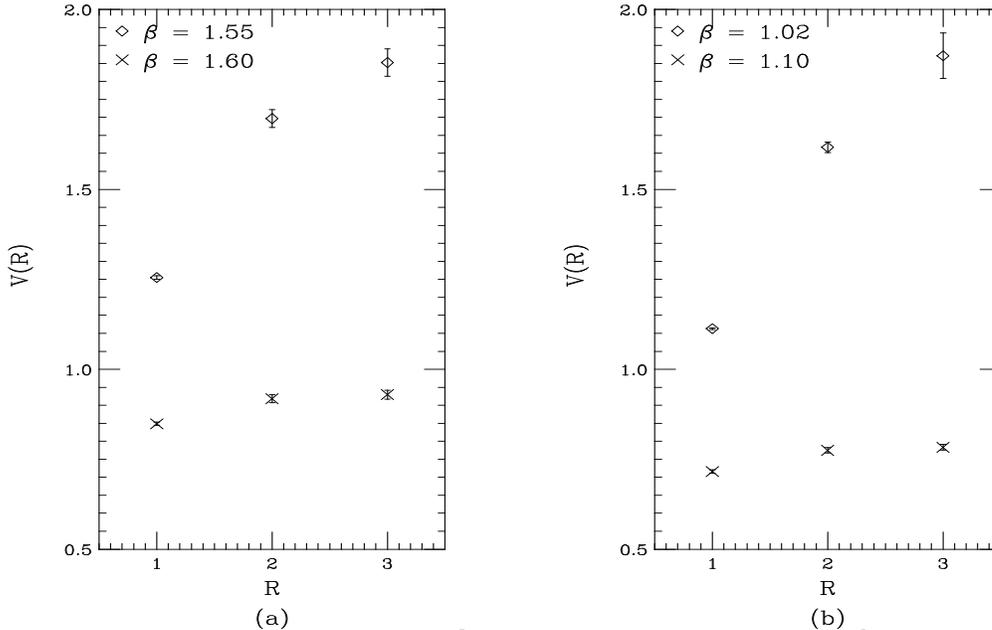

Fig. 1. Static quark potentials on a $6^3 \times 4$ fluctuating lattice with $m_p^2 = 0.005$ (a) and for a flat simplicial lattice with gravity switched off (b). One clearly sees the transition from the confined to screened behavior. Error bars arise from a jack-knife procedure.

To extract a reliable value for the string tension, we simulated the system on an $8^3 \times 4$ lattice. The results are shown in Fig. 2 where the solid lines are fits to the correlations in Eq. (3) according to the potentials of Eq. (4) with a mirror term included. Our data points lie on top of those obtained from the $6^3 \times 4$ lattice except for $R = 3$ which feels the mirror charges. On the fluctuating lattice, a value of the string tension $\sigma_{grav}^{simp} = 0.45(1)\langle x_l \rangle_s^{-1} \langle x_l \rangle_t^{-1}$ is extracted setting $\alpha = 0$. The average link along the main axes has a value of $\langle x_l \rangle_s \approx \langle x_l \rangle_t \approx 3.3 l_0$. Our data in the confined phase resolve practically no Coulomb effects in the potential between the quark sources, $\alpha = 0.07(15)$. For the pure gauge case, a string tension value of $\sigma_{flat}^{simp} = 0.42(2) a^{-2}$ is obtained with $\alpha = 0.20(4)$. The lattice spacing $a$ on the flat lattice can also be expressed in units of $l_0$, $a \approx 2.2 l_0$. We further fitted the Debye screened potentials in the deconfined phase. For the fluctuating lattice, we obtain a value of $\mu = 0.49(7)\langle x_l \rangle_s^{-1}$ with $\alpha = 0.24(1)$ while for the flat lattice, we find $\mu = 0.53(9) a^{-1}$ with $\alpha = 0.25(1)$.

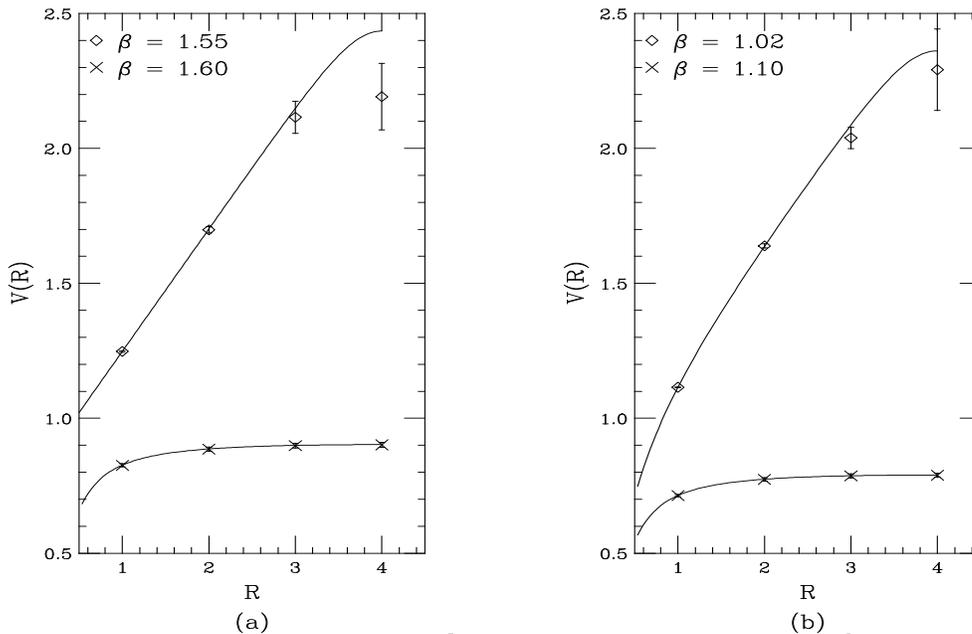

Fig. 2. Static quark potentials on an $8^3 \times 4$ fluctuating lattice with $m_p^2 = 0.005$ (a) and for a flat simplicial lattice with gravity switched off (b). The curves are fits to a confined and Debye screened potential with periodicity taken into account.

Table 1
Values of the Coulomb parameter $\alpha$, string tension $\sigma$, Debye screening mass $\mu$ and constant $C$ obtained from fits to the potential for the $8^3 \times 4$ lattices.

| Lattice | Confined | | | | Deconfined | | | |
| --- | --- | --- | --- | --- | --- | --- | --- | --- |
| | $\beta$ | $\alpha$ | $\sigma$ | $C$ | $\beta$ | $\alpha$ | $\mu$ | $C$ |
| Fluctuating | 1.55 | 0 | 0.45(1) | 0.79(1) | 1.60 | 0.24(1) | 0.49(7) | 1.08(1) |
| Flat | 1.02 | 0.20(4) | 0.42(2) | 0.89(6) | 1.10 | 0.25(1) | 0.53(9) | 0.97(1) |

Our results are summarized in Table 1.

Finally, we comment on the distance $R$ between quark sources on a fluctuating skeleton. The correct distance between two points should be measured using geodesic distances. The geometry is known to be Euclidean only inside the pentahedron with the curvature on the triangles. We take the distance $R$ between the source and sink to be equal to the index distance along the main axes of the skeleton. This seems a reasonable approximation in the well-defined phase with small curvature fluctuations. Other proposals using the methods of scalar field propagation [14] require corrections at small distances and need further exploration to be suitable for our purposes.

## 4  Conclusion

We have studied static SU(2) quark potentials in the presence of quantum gravity and find that confinement from the non-abelian gauge fields is not destroyed by quantum gravitational effects. In the confined phase we get a potential linearly rising with $R$, and in the deconfined phase we see a behavior resembling a screened Coulomb potential. We have extracted string tension values in the confined phase and screening masses in the deconfined phase for both the coupled system of quantum gravity with SU(2) gauge fields and for the pure gauge theory on a simplicial lattice without gravity. We have noted that the "entropy dominated" phase, $0 \leq m_p < m_c$, seems to be stable with the increase in $N_t$. A decrease in the ratio $T_c/m_c$ by a factor two was observed. To present evidence for a further fall-off in this ratio would require simulations on lattices with larger $N_t$ extensions. Another interesting possibility would be to study mass ratios $m_h/m_c$ on lattices with zero temperature.